\RequirePackage{fix-cm}
\documentclass[twocolumn]{svjour3}          
\smartqed  
\usepackage{graphicx}
\usepackage{psfrag}
\usepackage{pstricks,pst-grad}
\usepackage{xspace,colortbl}
\usepackage{amsmath}
\usepackage{amsfonts}
\usepackage{amssymb}

\newcommand{\vu}{\mbox{\boldmath$u$}}
\newcommand{\vv}{\mbox{\boldmath$v$}}

\newcommand{\vx}{\mbox{\boldmath$x$}}
\newcommand{\vy}{\mbox{\boldmath$y$}}

\newcommand{\vA}{\mbox{\boldmath$A$}}
\newcommand{\vB}{\mbox{\boldmath$B$}}
\newcommand{\vC}{\mbox{\boldmath$C$}}

\newcommand{\vE}{\mbox{\boldmath$E$}}
\newcommand{\vF}{\mbox{\boldmath$F$}}

\newcommand{\vI}{\mbox{\boldmath$I$}}
\newcommand{\vK}{\mbox{\boldmath$K$}}
\newcommand{\vP}{\mbox{\boldmath$P$}}
\newcommand{\vQ}{\mbox{\boldmath$Q$}}
\newcommand{\vR}{\mbox{\boldmath$R$}}
\newcommand{\vS}{\mbox{\boldmath$S$}}
\newcommand{\vT}{\mbox{\boldmath$T$}}

\newcommand{\vW}{\mbox{\boldmath$W$}}
\newcommand{\vX}{\mbox{\boldmath$X$}}
\newcommand{\vY}{\mbox{\boldmath$Y$}}

\newcommand{\vvarepsilon}{\mbox{\boldmath$\varepsilon$}}

\newcommand{\vGamma}{\mbox{\boldmath$\Gamma$}}
\newcommand{\vSigma}{\mbox{\boldmath$\Sigma$}}

\newcommand{\Z}{\mathbb{Z}}
\newcommand{\trace}{\mbox{Tr}}

\begin{document}

\title{A non-parametric efficient evaluation of Partial Directed Coherence\thanks{P.O. Amblard is supported by  a Marie Curie International Outgoing Fellowship from the European Union. Part of this work was performed while P.O. A. was affiliated with  the  University of Melbourne,  Math\&Stat Dept, Australia.}
}

\author{Pierre-Olivier Amblard   
}


\institute{P.O.A. \at
             GIPSAlab/CNRS UMR 5216, Grenoble, France\\
              Tel.: +33-476826358\\
              Fax: +33-476574790\\
              \email{bidou.amblard@gipsa-lab.inpg.fr}           
           }

\date{Received: date / Accepted: date}

\maketitle

\begin{abstract}
Studying the flow of information between different areas of the brain can be performed by using the so-called Partial Directed Coherence.
This measure is usually evaluated by first identifying a multivariate autoregressive model, and then by using Fourier transforms of the impulse responses identified and applying appropriate normalizations. Here, we present another route to evaluate the partial directed coherences in multivariate time series. The method proposed is non parametric, and utilises the strong spectral factorization of the inverse of the spectral density matrix of the multivariate process. To perform the factorization, we have recourse to an algorithm developed by Davis and his collaborators.  We present simulations as well as an application on a real data set (Local Field Potentials in the sleeping mouse)  to illustrate the methodology. A comparison to the usual approach in term of complexity is detailed. For long AR models, the proposed approach is of interest.

\end{abstract}
\sloppy
\section{Introduction}
\label{intro}

The last decade has seen a dramatic increase in the number of studies about connectivity in the brain \cite{Spor10}.
Important questions concern the modification of the connectivity in networks of the brain during development of illnesses. The problem of discovering connectivity from monitoring of the brain activity is therefore of crucial  importance. This problem is an inference problem. 
It can be given an elegant formulation using graph theory and the notion of graphical modeling of multivariate processes. Basically, a measurement ({\it e.g. } BOLD signal as measured by fMRI in a part of small zone of the brain, Local Field Potential (LFP) as delivered by an intracranial electrode, \ldots)
is associated to a node in a graph. The measurement of many different signals ({\it e.g. } many  cells in  fMRI, many LFP \ldots) thus defines the set of nodes of a graph. Inferring connectivity amounts to identifying the edges of the graph, based on the knowledge of the nodes. The edges can be undirected or directed. In the so-called functional connectivity \cite{Spor10}, an undirected link exists between two nodes if the corresponding measurements are sufficiently dependent. 

In this paper, we deal with directed edges. An appealing method to assess directional dependence between nodes is to use the notion of Granger causality, a concept now widely used in fields as diverse as economy, meteorology or neuroscience \cite{Gran80,GourBF06,Chic11}.
Granger causality states that a measurement is a cause of a signal if it helps in the prediction of this signal. This notion is relative to the set of measurements used. This means that adding a new measurement to the set may alter the conclusion drawn from the only set. This also implies  that when testing Granger causality between two measurements of a set, extra measurements of the set must be taken into account. Finally, Granger causality can be symmetrical: the fact that one signal causes a second one does not preclude the fact that the second signal causes the first one.  A nice development of graphical models based on Granger causality has been done by  \cite{Eich11}. 

In the preceding discussion, the notions of dependence and of predictions have remained vague. Strong definitions can be given in probabilistic terms. For example, Granger causality can be defined using concepts of conditional independence \cite{Eich11}, leading to practical measures based on directed information theory \cite{AmblM11,AmblM13}. But weaker definitions also exists that allows efficient and robust practical inference algorithms to be implemented. Among the weaker definition, those relying on linear modeling and Gaussianity are appealing, since almost all the theory of practical inference can be explicitly written down \cite{Gewe82,Gewe84}. Furthermore, linear modeling paves the way to a possible frequency domain interpretation of Granger causality. Among the different tools that has been developed, Partial Directed Coherence (PDC) has an important place in the landscape. It is now a well accepted tool in neuroscience to assess direction of information flow between different areas of the brain \cite{SameB99,BaccS01,ScheWEPHGLDT05,ScheTE09}. It relies in identifying links between
two areas of the brain using causal linear filters. The technique can be applied to any kind of multivariate measurements, as soon as
the measured signals are jointly stationary times series. For example, this can be applied to monitor the flow of energy between different areas of the brain using fMRI measurements, or can reveal the circulation of information between electrode measuring 
local field potentials. In this paper, the illustration of the technique we propose concerns local field potentials
recorded simultaneously {\it in-vivo} in mice brains. 

The interpretation of PDC in the Granger causality framework relies on the Wold decomposition of multivariate signals, meaning that a second-order stationary process can be viewed as the output of a multivariate linear systems attacked by a multivariate white noise. In most cases, the decomposition is invertible, and the process can be view as an infinite horizon autoregression attacked by the white noise. It is in this case that Geweke's indices for Granger causality  make sense \cite{Gewe82,Gewe84}, and it is also in this case that PDC can be viewed as a way of assessing Granger causality in the frequency domain. 

Up to now however, the practical evaluation of PDC relies on a finite horizon autoregression modeling, and uses the toolbox of multivariate AR modeling \cite{BaccS01,Lutk05}. This leads to the usual problems of parametric modeling, the more important being the order determination. Another problem may occur for large scale studies, for which a large number of signals is at hand, and if large orders are needed. In this case, the usual determination of the AR models may require inversion of very large matrices leading to impossible calculation due to  heavy computational burden. 

We propose here a direct evaluation of transfer functions between any pair of the measured signals, and hence to a direct evaluation of PDC for any pair of signals. It is based on the strong spectral factorization of the inverse of the spectral matrix of the signals. The method is rapid and non parametric in nature, and thus allows a full interpretation of Granger causality in Geweke's sense.

During the course of the work, we discovered the work by   \cite{DhamRD08} where the idea of explicit spectral factorization is also used. However, it is used on 2$\times$2 matrices only to study the transfer between a pair of signals. The transfer is then used to calculate a Geweke's index in the frequency domain. This is then repeated for all possible pairs of signals.  In contrast, our work deals with the whole spectral matrix. Precisely, we factorize the inverse of the spectral matrix thus leading to the whole hierarchy of transfer functions between any pair of signals conditionally to the others. Another difference with the work in   \cite{DhamRD08} is the use of another spectral factorization algorithm. In   \cite{DhamRD08}, Wilson's algorithm is used \cite{Wils72}. We prefer to use Dickinson\&Davis algorithm developed in 1978 \cite{Dick78,DaviD83}. The  latter is firmly grounded on causal filtering principles, whereas the former came from an {\it ad-hoc} application of a Newton-Raphson iteration. However, these two algorithms are quite similar, use a causal projection operator, and more importantly satisfies quadratic convergence (as issued from Newton-Raphson iterations). For the sake of completeness, we will recall in an  appendix the derivation of Dickinson\&Davis. Furthermore, we provide the Matlab/Octave code for the spectral factorization algorithm.

The paper is organized as follows. The main section is devoted to the presentation of the method and to its application to a synthetic example as well  as real recordings of local fields potentials in the brain of a sleeping mouse. We will insist in the course of the presentation on the importance of the Wold decomposition in the interpretation we may have of the PDC. Practicalities concerning the spectral factorization algorithm as well as some statistical issues will be developed. In the last section, we will discuss advantages and drawbacks of the approach, and will provide a detailed comparison with the usual method (multivariate AR modeling) in term of complexity analysis.

\section{From Wold decomposition to  Partial Directed Coherence {\it via } spectral factorization}
\label{sec:1}

Consider a neuroscience experiment where $p$ signals are simultaneously recorded. This can be a functional Magnetic Resonance Imaging experiment, during which the brain of a subject is monitored while doing a task; this can be a MEG or an EEG recording session; 
this can be the monitoring of an animal equipped with intracranial multi-electrode devices; {\it etc}. 
For this experiment, we store the $p$ simultaneous  measurements  into a multivariate process $\vx(t)$ of dimension $p$. 
In the following, $^\top$ stands for the transposition of a vector or a matrix, $\vI$ stands for the identity matrix of appropriate dimension, and recall that $ \vA \leq \vB$ for matrices is understood as $\vB-\vA$ is positive definite.

\subsection{Wold decomposition and the linear model}

We now assume that the multivariate process can be mathematically described by a second-order stationary multivariate stochastic process 
$\vx(t)$, where $t$ is a discrete time parameter.  The Wold decomposition \cite{Roza67} then states that this process can be represented as
\begin{eqnarray*}
\vx(t) = \sum_{k \geq 0} \vB(k)  \vvarepsilon(t-k)
\end{eqnarray*}
where for each $k\geq 0$, $\vB$ is a matrix of  size $p\times p$, and where $\vvarepsilon(t)$ is a multivariate zero mean white noise process of dimension $p$, with covariance matrix $\Sigma_{\varepsilon} $. Precisely, if $E[.]$ denotes the mathematical expectation operator (ensemble average), we have $E[ \vvarepsilon(t)]=0$ and $E[\vvarepsilon(t)\vvarepsilon^\top(t+\tau)] = \Sigma_{\varepsilon}   \delta_{t,\tau}$.

Then $\vx(t)$ has a spectral density matrix  $\vS_{xx}(\lambda) = \sum_{k\in \Z} E[\vx(t)\vx^\top(t+\tau)] \exp(-2 \iota \pi \lambda t)$.
If the spectral density matrix is bounded and strictly positive definite in the sense that $c_1 \vI  \leq \vS_{xx}(\lambda)  \leq c_2 \vI$ for certain constants $0< c_1\leq c_2 < +\infty$, then it is possible to invert the Wold decomposition and write
\begin{eqnarray}
\vx(t)  = \sum_{k=1}^{+\infty} \vA(k) \vx(t-k)   + \vvarepsilon(t)
\label{ARinfini:eq}
\end{eqnarray}
where for each $k$, $\vA$ is a matrix of  size $p\times p$.  To understand this model,
consider the $j$-th component $x_j(t)$ and write down its full expression as a function all the components $x_i$.
We have
\begin{eqnarray}
x_j(t) &= & \sum_{i=1}^p x_{i\rightarrow j}(t)+ \varepsilon_j(t)  \\
 x_{i\rightarrow j}(t)&=&\sum_{k=1}^{+\infty} A_{ji}(k) x_i(t-k) 
 \label{modele:eq}
\end{eqnarray}
Thus $x_j$ at time $t$ is modeled as the sum of the influence of its past on itself with the influences of the past of the other components on itself. Here, the influences are modeled with linear links.  The term $\varepsilon_i(t)$ is the innovation sequence of the process $i$. $\vvarepsilon(t)$ is a multivariate white noise sequence, in the sense that two samples at different times are uncorrelated.

Equation (\ref{ARinfini:eq}) is a very general mathematical representation for the multivariate signal, and as developed by \cite{Gewe82,Gewe84}, this is the strict framework in which Granger causality has a firm meaning. We insist on this by making some remarks:
\begin{itemize}
\item The only requirements for it to be valid have been recalled : the process should be a second-order stationary process (meaning that $\trace( E[\vx(t)\vx(t)^\top] ) < +\infty$, constant mean and $E[\vx(t) \vx(t+h)^\top]$ is a function of $h$ only). No assumpion of Gaussianity  is made.
\item Furthermore, the Wold decomposition is a representation of the process and must not be considered as a physical model of it. To insist on this, take the example of a signal $x(t)$ obtained by nonlinear transforming another one. If $x(t)$ is second-order stationary, it will admit the linear Wold representation for some innovation sequence $\vvarepsilon(t)$ and some sequence of matrices $\vB(k)$. If inversion of the model is possible, equation {\ref{ARinfini:eq}} will be satisfied.
Therefore in general, this model should not be interpreted as a model describing the physics of the interaction between different parts of the brain, or if it is, it should be  {\em  only with caution}.

\item The representation precludes the use of correlated noise in the model, as used for example to represent exogeneous inputs. 
\end{itemize}
All this written, the representation (\ref{ARinfini:eq}) is often manipulated as if it was the physical reality that produced $\vx(t)$. We also do this in the following but knowing the caution recalled.

The summations in (\ref{modele:eq}) begins with $k=1$. $k=0$ could also be included to model an instantaneous link between the variables. This  could practically exist: for example, any dynamical interaction between two signals that occurs more rapidly than the sampling period will be perceived as an instantaneous interaction between them. However, if the summation starts with $k=0$ the model suffers a problem of identifiability. To eliminate this problem, it is possible to reject the instantaneous interaction into the dynamical noise 
$\vvarepsilon(t)$: the correlation between the components of this multivariate noise models the possible instantaneous interactions between the signal components (see \cite{Gewe82} for example).

The model is a particular instance of Granger causality graphs introduced by  \cite{Eich11}. Granger causality graphs are graphical models of multivariate times series. A node in the graph represents one component of the multivariate signal. Here, node $i$ will represent
signal $x_i(t)$. A directed edge from node $j$ to node $i$ exists if and only if signal $x_j$ Granger causes signal $x_i$ (conditionally to the other signals), which in the case of the model considered in this paper is equivalent to $A_{ji}(k)$ is not identically zero. Testing for Granger causality in the model of this paper,  and thus testing for the possible influence of the past of one signal onto another, is thus equivalent to testing the non nullity of an impulse response. Equivalently, we can study the so-called transfer function which is nothing but the Fourier transform of the impulse response. Therefore, a fundamental problem here is to identify the impulse responses $A_{ji}(k), k>0$, or equivalently their Fourier or their $z$ transforms.

\subsection{Identification of the model}

Usually, the model is  identified from data using least square methods. To perform the identification practically, the time horizon of the impulse responses is considered finite. In other words, the multivariate process is supposed to be Markovian. Then, the matrices $\vA(k)$ are identified using tools from multivariate autoregressive modeling. The methods are inherently parametric. They include the choice of the maximal time horizon in the past. Indeed, $ x_{j\rightarrow i}(t)$ is in general   modeled as
\begin{eqnarray*}
x_{j\rightarrow i}(t) = \sum_{k=1}^{q_{ji}}A_{ji}(k) x_i(t-k) + \varepsilon_i(t) 
\end{eqnarray*}
and the inference procedure not only concentrates on the impulse responses $A_{ji}(k)$ but also on the orders $q_{ji}$. In general, identification methods use a mean square error approach coupled with model order selection criterion (such as BIC or AIC, or others) \cite{Lutk05}.
If the orders are all the same (we assume this for the sake of simplicity) and equal to $q$, the usual identification methods use vectors of size $q p$ and matrices of size $(qp\times qp)$ which can be very large, leading to heavy computational burden. 

Note again that this approach departs from the original interpretation of the model as the inversion of the Wold decomposition.
Here, we will stick more closely to the original model without imposing a finite time horizon (other than that imposed by the finite length of the data). The method we adopt is then inherently non parametric and deals closely with the original equation (\ref{ARinfini:eq}). Furthermore, as we will described shortly, the analysis sticks with the well-known analysis of graphical modeling of multivariate variable in statistics \cite{Whit89}

The advantage of the method is twofold. Firstly, as a non parametric methodology, we are not stuck with the problem of order selection and we do not suffer of any assumption on the models. Secondly, the algorithm relies on a very efficient algorithm for strong spectral factorization which is very fast.

In the following, we work with the $z$ transform, defined for a function $y(t)$ as $Y(z) = \sum_{t\geq 0} y(t) z^{-t}$. The sum is assumed to be convergent. For functions that grows to infinity at most exponentially fast, this requires that the complex number $z$ lies in some disk centered at the origin. We will assume that  the unit circle is included in that disk. For $z=\exp(2\iota \pi \lambda)$ on the unit circle ($\iota^2=-1$), we obtain the discrete time Fourier transform $Y(\lambda)$ of $y$ (note the abuse of notation $Y(\lambda)=Y(z=\exp(2\iota \pi \lambda))$).  When working with matrices, the transforms are taken component wise. We will denote by $z^\star$ the complex conjugate of $z$, by $z^{-\star}$ the complex conjugate of $z^{-1}$, by $\vA^\dagger$ the Hermitian transpose of  the matrix  $\vA$, and by $\vA^\top$ its usual transpose. $\vI$ stands for the identity matrix of adequate dimension. 
 
Since the noise $\vvarepsilon(t)$ is a white sequence, the multivariate process $\vx$ admits the following 
spectral density matrix
\begin{eqnarray*}
\vS_{xx}(z) = (\vI - \vA(z) )^{-1} \Sigma_{\varepsilon} (\vI - \vA(z^{-\star}) )^{-\dagger}
\end{eqnarray*}
where $\vA(z)$ is the matrix of the $z$ transform (element wise) of the sequence of  matrices $\vA(k)$.  Therefore we get 
\begin{eqnarray}
\vS_{xx}(z)^{-1} =  (\vI - \vA(z^{-\star}) )^{\dagger} \Sigma_{\varepsilon}^{-1} (\vI - \vA(z) )
\label{invSpecMat:eq}
\end{eqnarray}

Consider now the problem of strong spectral factorization. This problem occurs in optimal linear filtering and control theory \cite{AndeM79}, when the need of causal filters or controllers  is required. Solving optimal causal linear filtering in the multivariate case requires to solve the spectral factorization of the spectral density matrix of the observation process, say $\vS_{xx}(z) $. This matrix is Hermitian, positive-definite and is defined as the $z$ transform of the correlation matrix $E[\vx(t) \vx^\dagger(t-k)]$. As such, it admits a strong factorization 
\begin{eqnarray}
\vS_{xx}(z) = \vF(z)  \vW \vF^\dagger(z^{-\star}) 
\label{stronfactor:eq}
\end{eqnarray}
 where $\vF(z) = \sum_{k\geq 0} \vF(k) z^{-k} $ is the $z$ transform
of a causal sequence of matrices and where $\vW$ is a positive definite matrix. Furthermore, $\vF(z)$ is invertible and its inverse is also the $z$ transform of a causal sequence of matrices. Then, the inverse of the spectral matrix also admits a strong factorization, with $\vF(z)^{-1}$ and $\vW^{-1}$ as spectral factors.

Comparing the result (\ref{invSpecMat:eq}) to the factorization  (\ref{stronfactor:eq}) we conclude that  the factor $\vI - \vA(z) $ is the strong spectral factor of the inverse of the spectral matrix.
Therefore, we have a way to identify the model (\ref{modele:eq}) from data: it suffices to estimate the spectral matrix from these data and to perform the spectral factorization of the inverse of this matrix to obtain an estimate of $\vI - \vA(z)$.

 \subsection{Spectral factorization algorithm}
 
 \paragraph{Definition.}
 
We use the factorization algorithm designed by  J. H. Davis and his collaborators \cite{Dick78,DaviD83,HarrD92}. This algorithm is iterative and from an  initial guess $\vF_0(z)$ 
builds up the sequence for $n\geq 0$
\begin{eqnarray*}
\vW_n &=& \frac{1}{2\iota \pi}\oint \vF_n(z)^{-\dagger} \vS_{xx}(z)^{-1} \vF_n(z)^{-1}  \frac{dz}{z} \\
\vF_{n+1}(z)  &=& \vW_n^{-1} P_+ \Big( \vF_n(z)^{-\dagger} \vS_{xx}(z)^{-1} \vF_n(z)^{-1} \Big) \vF_n(z) 
\end{eqnarray*}
The operator $P_+$ is the causal projection operator. For a $z$ transform $H(z)$ of a bilateral sequence $h_k$ it is defined as 
\begin{eqnarray*}
(P_+H)(z) = \sum_{k\geq 0} h_k z^{-k}  =  \frac{1}{2i\pi }\oint  \frac{dv}{v} \frac{H(v)}{1-vz^{-1}} 
\end{eqnarray*}
It simply consists in truncating the domain over which the $z$ transform is calculated. 
It was shown by  \cite{Dick78}  that  the iterated $\vF_{n}(z) $ converges 
almost everywhere to $\vF(z)$, and that $\vW_n$ converges to $\vW$. Of course, this is valid under some technicalities, among which 
$\vF_0$ should be the $z$-transform of a causal sequence of matrices. Practically, initializing these matrices to be the identity is sufficient. We give some details on the derivation of this algorithm in appendix \ref{algoDickinson:app}. As is recalled there, the derivation relies on a Newton-Raphson iteration applied to a Riccatti equation. The algorithm then inherits of the well-known fast quadratic convergence rate of Newton-Raphson algorithms \cite{BoydV04}

Practically, we will work of course with real frequency rather than complex variables $z$. And furthermore, since we are dealing with finite size data, we will end up with discrete frequencies.

\paragraph{Practicalities.}
If we work with data sampled at the frequency $f_s$, on signals of length $N$, then we will consider the discrete frequencies $m f_s /N$, with $m$ varying from $-N/2+1$ to $N/2$. Since spectral matrices are Hermitian, the positive frequencies are enough for a complete description.

$P_+$ is the projector over the space of matrices  with  entries which are Fourier transform of causal sequences. $(P_+X)(m)$ is implemented using the inverse Discrete Fourier Transform (DFT). The idea is to invert the DFT to obtain $x_n, n=0,\ldots,N-1$, multiply by a step function to set to zero the values of the function at negative times, and to transform back. However, the step function must be chosen in order to respect symmetries and the periodicity of the DFT. 
Recall that for real signals, these properties implies that  the $N$ samples of the signals correspond to one period. Thus in general, the first $N/2+1$ represent the positive times whereas the $N/2-1$ remaining represents the strictly negative times. Let $u_n$ be the step function used to keep the causal part, {\it i.e. } the positive times. A naive choice would be to set $u_n=1, \forall n=0,\ldots, N/2$ and $u_n=0, \forall n=N/2+1,\ldots, N-1$. Doing so violates the symmetries mentioned above. To satisfy these symmetries, we introduce $v_n$ the step function used to select the anti-causal part of a sequence.  We can write explicitly
   \begin{eqnarray*}
(P_+X)(m) & = &\sum_{n=0}^{N-1} u_n x_n e^{-2\iota n m /N} \\
X_-(m) & = &\sum_{n=0}^{N-1} v_n x_n e^{-2\iota n m /N} \\
\end{eqnarray*}
Then, if we consider the decomposition of $X(m)$ as the sum of the causal part $(P_+X)(m)= X_+(m)$ and the anti causal part $X_-(m)$, we must have $X^\star_+(m)=X_-(m)$,
and  necessarily $u_{N-n}=v_n, \forall n=0,\ldots, N-1$. In particular, if $N$ is even (which is practically true if we use the Fast Fourier Transform), we must have $u_{N/2}=v_{N/2}$. Therefore, the step function $u_n$ must be chosen as 
\begin{eqnarray*}
u_n= \left\{ 
 \begin{array}{lcl}
1 & \mbox{ if}& n=0,\ldots,N/2-1\\
1/2 &\mbox{ if}& n=N/2\\
0&\mbox{ if}& n=N/2+1,\ldots,N-1\\
\end{array} \right.
\end{eqnarray*}
The 1/2 term can be understood as a consequence of the periodicity induced by the use of the DFT.

To apply the algorithm, we first have to estimate $\vS_{xx}(m), \forall m=0,\ldots, N-1$. This can be done using any standard non parametric spectral estimation algorithm. If the length of the data is small, a nice possibility is to use multitaper spectral estimation \cite{PercW93}, or smoothing of the periodogram \cite{Bril01}. Here, however, since we will apply the algorithm to long data, we use the averaged periodogram method, also known as Welsh method. Basically, the signal is cut into $K$ blocks of size $N$, each block is Fourier transformed, then squared, and the estimated spectrum evaluated by averaging over the blocks. This is done also for the cross-spectra. In short, we use the estimator
\begin{eqnarray}
\widehat{\vS}_{xx}(m)& =& \frac{1}{K}  \sum_{k=0}^{K-1} \frac{1}{N}\vX_k(m) \vX_k(m)^\dagger   \label{estimator:eq} \\
\vX_k(m) &=& \sum_{n=0}^{N-1} \vx(n+k N) h_n e^{-2\iota \pi \frac{m n}{N}} 
\end{eqnarray} 
where $h_n$ is an optional  tapering window (as the Hamming window) of unit energy ($\sum_{n=0}^{N-1}h_n^2=1$). The term $1/N$ in (\ref{estimator:eq}) is necessary to ensure convergence in the mean of 
$ \frac{1}{N}\vX_k(m) \vX_k(m)^\dagger$ to the true value $\vS_{xx}(m)$ as the size $N$ of the blocks tends to infinity.

The following algorithm is then applied to the estimated spectral matrix. 
In discrete frequency the algorithm reads
\begin{eqnarray*}
\vF_{n+1}(m)  &=& \vW_n^{-1} P_+ \Big( \vF_n(m)^{-\dagger} \widehat{\vS}_{xx}(m)^{-1} \vF_n(m)^{-1} \Big) \vF_n(m) \\
\vW_n &=&\sum_{m=0}^{N-1} \vF_n(m)^{-\dagger} \widehat{\vS}_{xx}(m)^{-1} \vF_n(m)^{-1}
\end{eqnarray*}
The algorithm is iterated until the norm of $\widehat{\vS}_{xx}(m)^{-1} - \vF(m)^{\dagger} \vW \vF_n(m) $ is lower than some prescribed tolerance. We give the full code for the algorithm in appendix \ref{Code:app}.

\subsection{Exploiting the spectral factors}

When the spectral factors are obtained, it remains to use them to practically assess flows of information.
Recall that $\vF(m)=\vI - \vA(m)$, and thus we get for $i \not= j$, $A_{ij}(m)=-F_{ij}(m)$.

We can use this to evaluate the Partial Directed Coherence (PDC), as defined by  \cite{BaccS01},
\begin{eqnarray*}
P_{j\rightarrow i}(m) = \frac{\left| A_{ij}(m)\right|}{\sqrt{\sum_k \left| A_{kj}(m)\right|^2}   }
\end{eqnarray*}
The PDC quantifies at each frequency bin $m$ the linear influence of signal $j$ onto $i$ as compared to the influence of $j$ onto all the other signals. The normalization adopted enforces  $P_{j\rightarrow i}(m) $ to be lower than 1.

As discussed in {\it e.g. } \cite{ScheTE09}, this normalization is however arbitrary, and the definition of the PDC suffers from some drawbacks. The main drawback is certainly its non invariance with respect to scales, which can be 
an important problem when dealing with signals measured in different units. Furthermore, the second order statistics of $P_{j\rightarrow i}(m) $ depends on the frequency. To circumvent these problems, a different normalization is introduced in \cite{ScheTE09}, which is statistical in nature, but which solve the problems raised.

In fact, a definition of PDC is valid if  signal $j$ does not influence $i$ is equivalent to $P_{j\rightarrow i}(m) =0, \forall m$. We thus see that the fundamental point is that the PDC should be proportional to $\left| A_{ij}(m)\right|$:
The real test of linear  influence is indeed whether $A_{ij}(m)$ is zero or not!
Hence, we should use $\left| A_{ij}(m)\right|$ as a test statistics. In order to get good statistical properties, it is natural to normalize this statistics by is variance!

When the model is identified by least square fitting of a multivariate model, explicit asymptotic results can be obtained for the variance \cite{ScheTE09}. This however depends on the true parameters, and their estimates have to be used. 

In our case, we do not have yet this expression. Obviously the statistics of the estimate $\widehat{\vS}_{xx}(m)^{-1}$ are known asymptotically in the size of the blocks $N$, because $\widehat{\vS}_{xx}(m)$ can be shown to be 
asymptotically (in $N$) distributed as a Wishart random matrix under mild assumption on the process (mixing conditions) \cite{Bril01}. Thus  $\widehat{\vS}_{xx}(m)^{-1}$  is asymptotically an inverse Wishart, from which its statistics can be computed. For example, it can be shown that it is asymptotically unbiased in the number  $K$ of blocks (when
in fact $\widehat{\vS}_{xx}(m)$ is unbiased). Likewise, the variance of the elements of the matrix can be evaluated. However, we did not succeed in obtaining the statistics of the spectral factors from the statistics of 
$\widehat{\vS}_{xx}(m)^{-1}$. 

 But we can use the parametric bootstrap to estimate this variance \cite{Hall92}. When the spectral factors are estimated, we then get estimates for  $\vA(k)$ and $\Sigma_\varepsilon$, and we can generate data using this estimated model. Thus we can obtain a bootstrap estimate 
 $V_{ij}$ of the variance. This variance is use to normalize $\left| A_{ij}(m)\right|^2$ to define the statistics
 \begin{eqnarray*}
P_{j\rightarrow i}(m) = \frac{\left| A_{ij}(m)\right|^2}{V_{ij}   }
\end{eqnarray*}
which is, under the hypothesis of no influence,  asymptotically (in $K$) $\chi^2_2/2$ as the square of a (asymptotically in $K$) complex normalized normal random variable.   In fact, we must say that we conjecture this last result. The reasons for that conjecture are the following.
Under mild assumptions on the multivariate process (its correlation function should decrease fast enough to be summable), we already mentioned that the estimate of the spectral density matrix is asymptotically a complex circular Gaussian law at each frequency (circular mean independence between the real and the imaginary part), and that at two different frequencies, the estimates are independent. These two results remains exact for the inverse  of $\widehat{\vS}(m)$. The real conjecture is to suppose that the application which associates a spectral density matrix with the pair of its spectral factors $(\vF(m), \vW)$ is differentiable. If true, the delta-method can be applied to conclude that 
the estimated pair $(\vF(m), \vW)$ will converge to a complex Gaussian distribution. However, we cannot say if it is circular or not since we do not know the Jacobian of the application, and do not have access to a closed form expression of the covariance of this Gaussian.

We thus assume the conjecture, and we can then  set up a Family Wise Error Rate test of rate $\alpha$. Signal $x_j$ will be declared to have an influence over $x_i$ whenever $P_{j\rightarrow i}(m) >2 \eta(\alpha)/(N/2+1) $  for some $m$, $\eta(\alpha)$ being the $\alpha$-percentile of the chi square distribution with two degrees of freedom.  The $1/(N/2+1)$ factor corresponds to the well-known  Bonferroni correction to take into account the $N/2+1$ frequencies  tested \cite{Effr10}.

\subsection{A synthetic example}

\begin{figure}

\begin{picture}(0,0)%
\includegraphics{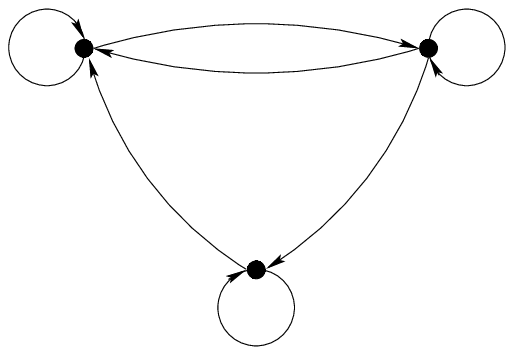}%
\end{picture}%
\setlength{\unitlength}{2072sp}%
\begingroup\makeatletter\ifx\SetFigFont\undefined%
\gdef\SetFigFont#1#2#3#4#5{%
  \reset@font\fontsize{#1}{#2pt}%
  \fontfamily{#3}\fontseries{#4}\fontshape{#5}%
  \selectfont}%
\fi\endgroup%
\begin{picture}(5079,3659)(1521,-2924)
\put(4201,-2183){\makebox(0,0)[lb]{\smash{{\SetFigFont{9}{10.8}{\rmdefault}{\mddefault}{\updefault}{\color[rgb]{0,0,0}$x_2$}%
}}}}
\put(2307,112){\makebox(0,0)[lb]{\smash{{\SetFigFont{9}{10.8}{\rmdefault}{\mddefault}{\updefault}{\color[rgb]{0,0,0}$x_1$}%
}}}}
\put(1536,114){\makebox(0,0)[lb]{\smash{{\SetFigFont{9}{10.8}{\rmdefault}{\mddefault}{\updefault}{\color[rgb]{0,0,0}$A_{11}$}%
}}}}
\put(4204,-2828){\makebox(0,0)[lb]{\smash{{\SetFigFont{9}{10.8}{\rmdefault}{\mddefault}{\updefault}{\color[rgb]{0,0,0}$A_{22}$}%
}}}}
\put(6585,106){\makebox(0,0)[lb]{\smash{{\SetFigFont{9}{10.8}{\rmdefault}{\mddefault}{\updefault}{\color[rgb]{0,0,0}$A_{33}$}%
}}}}
\put(4194,504){\makebox(0,0)[lb]{\smash{{\SetFigFont{9}{10.8}{\rmdefault}{\mddefault}{\updefault}{\color[rgb]{0,0,0}$A_{31}$}%
}}}}
\put(4201,-321){\makebox(0,0)[lb]{\smash{{\SetFigFont{9}{10.8}{\rmdefault}{\mddefault}{\updefault}{\color[rgb]{0,0,0}$A_{13}$}%
}}}}
\put(5361,-1146){\makebox(0,0)[lb]{\smash{{\SetFigFont{9}{10.8}{\rmdefault}{\mddefault}{\updefault}{\color[rgb]{0,0,0}$A_{23}$}%
}}}}
\put(2812,-1154){\makebox(0,0)[lb]{\smash{{\SetFigFont{9}{10.8}{\rmdefault}{\mddefault}{\updefault}{\color[rgb]{0,0,0}$A_{12}$}%
}}}}
\put(5964,103){\makebox(0,0)[lb]{\smash{{\SetFigFont{9}{10.8}{\rmdefault}{\mddefault}{\updefault}{\color[rgb]{0,0,0}$x_3$}%
}}}}
\end{picture}%

 \includegraphics[scale=.5]{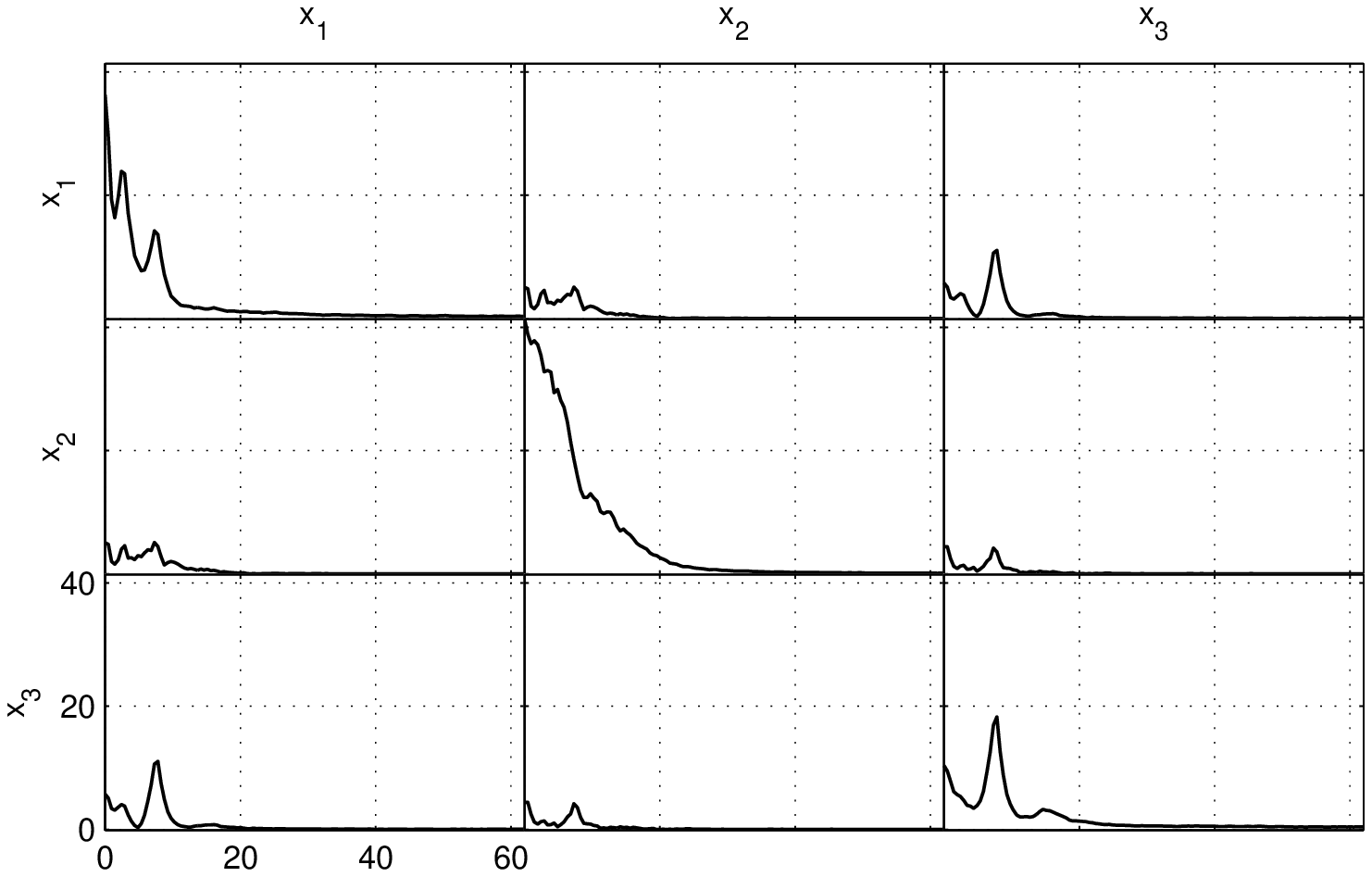}
  
    \includegraphics[scale=.5]{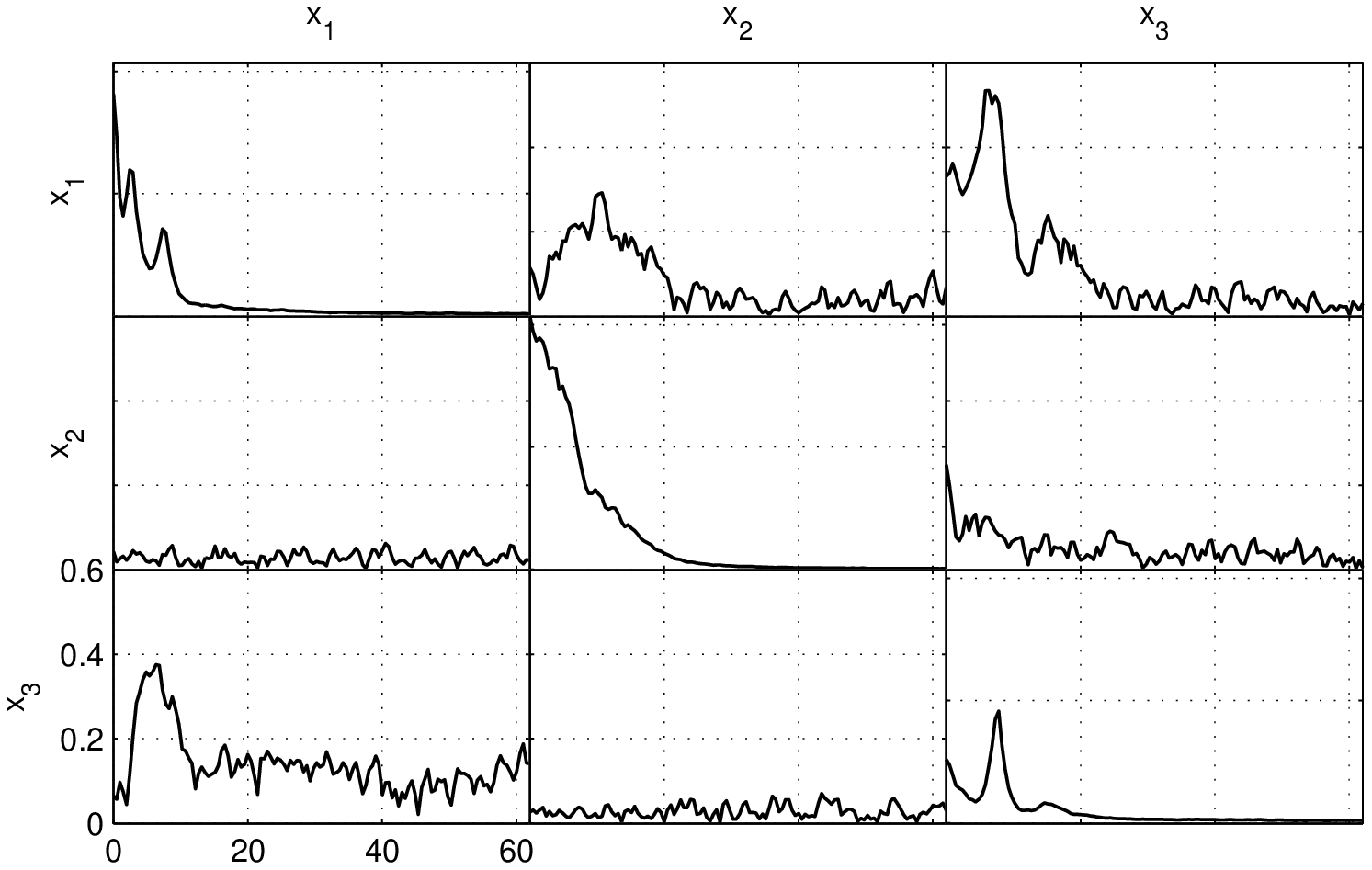} 
    
 \includegraphics[scale=.5]{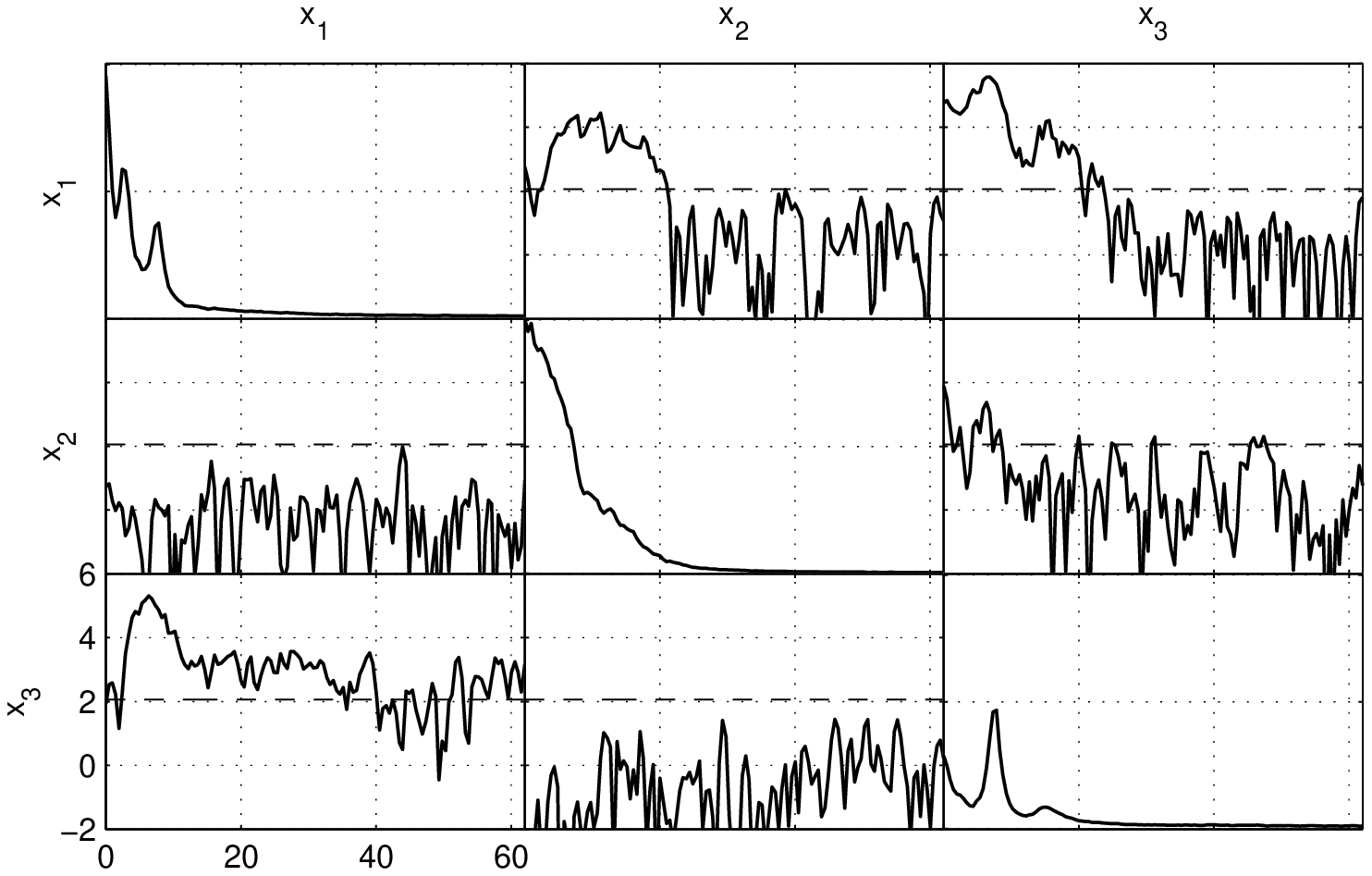}

\caption{Synthetic model used for illustration purposes. {\bf 1st row}: graphical model representation of the three dimensional signal studied.
{\bf 2nd row}: Spectral density matrix of the signal. All the subplots are displayed on the same linear scales as indicated in the lower left subplot.
{\bf 3rd row}: Directed Partial Covariance, except on the diagonal where the power spectral density of $x_i$ are plotted. All the subplots are displayed on the same linear scales as indicated in the lower left subplot, except the diagonal which are on the same $x$ scale, but on an arbitrary amplitude scale. {\bf 4th row}: Renormalized PDC, plotted on a log scale in amplitude.All the subplots are displayed on the same linear scales as indicated in the lower left subplot, except the diagonal which are on the same $x$ scale, but on an arbitrary amplitude scale. The dashed line correspond to the Bonferroni  threshold for a Family Wise Eror Rate of 0.05.  
For all the plots, the $x$ axis is labelled in frequency from 0 to 62.5 hz.  }
\label{ExempleSynth:fig}       
\end{figure}

To illustrate we consider here a three dimensional model
depicted in figure \ref{ExempleSynth:fig}.  To generate the model, we used real data in order to get realistic spectra. The data used are those described later in section \ref{RealData:sec}, and we therefore do not describe them yet. We chose three times series, identified a multivariate 
autoregressive model from them using a usual least square approach \cite{Lutk05} to obtain a sequence of matrices $\vA(k), k=1,\ldots, 50$.  
We then artificially set to zero the filters $A_{21}(k)$ and $A_{32}(k)$ in order to fit to the structure described in the figure.

The matrices $\vA(k)$ were then used to generate a synthetic time series using the equation 
\begin{eqnarray}
\vx(t)  = \sum_{k=1}^{50} \vA(k) \vx(t-k)   + \vvarepsilon(t)
\end{eqnarray}
where the white noise $\vvarepsilon(t)$ is chosen to have the identity as covariance matrix. We generated 566 blocks of length 256 samples,
and then applied the whole procedure. We show in figure \ref{ExempleSynth:fig} the spectral density matrix of the signal generated,
the PDC as calculated usually \cite{BaccS01} and the renormalized PDC \cite{ScheTE09} evaluated using the spectral factorization algorithm and the bootstrap variance estimation. Note that the renormalized PDC is depicted in log-scale in amplitude, and compared to the threshold corresponding to a  Family Wise Error Rate test of rate $\alpha$, using the Bonferonni correction. The threshold chosen is $\eta\left(\alpha (N/2+1)^{-1}\right)\approx7.8$ for $N=256$ and $\alpha=0.05$. The graphical model structure is correctly inferred from the renormalized PDC.

\subsection{Information flows between LFPs in the sleeping mice}
\label{RealData:sec}

We have access to data recorded in the sleeping mice during the paradoxical sleep phase.  The recording consists in intracranial local field potentials, with electrodes placed in several different areas of the brain (ParaFrontal Cortex--PFC--, Motor Cortex --M1--, Sensory cortex --S1,S2--, Ventral PosteroMedial nucleus  --vpm--,
and hippocampus --dCA1--). The position of the electrodes have been verified with a post experiment surgery. The aim here in analyzing the data is to show the effectiveness of the proposed method on real data. We do not intend to draw here any conclusion concerning the behavior of the brain.
The  application of this method in neuroscience experiments is under his way and will be presented elsewhere. 

\paragraph{Brief description of the data and parameter used.} The data consists in a six dimensional time series. It was recorded using a sampling frequency of 1000 Hz, using appropriate anti-aliasing filters. After inspection of the data, it appears that they are largely oversampled, and a digital under sampling by a factor of 8 is applied, leading to a new sampling frequency of  $ f_s=125 Hz$. 
At this rate, the length of the signal is of 145000 samples. 
We will present the application of the method to evaluate the flow of information between the six electrodes by means of the renormalized 
PDC. We apply the method at  frequency resolution: $\Delta m = f_s/ N$ with $N=256$.  The statistics is then
composed of $K= 566$ blocks.  

\begin{figure}
\includegraphics[scale=.55]{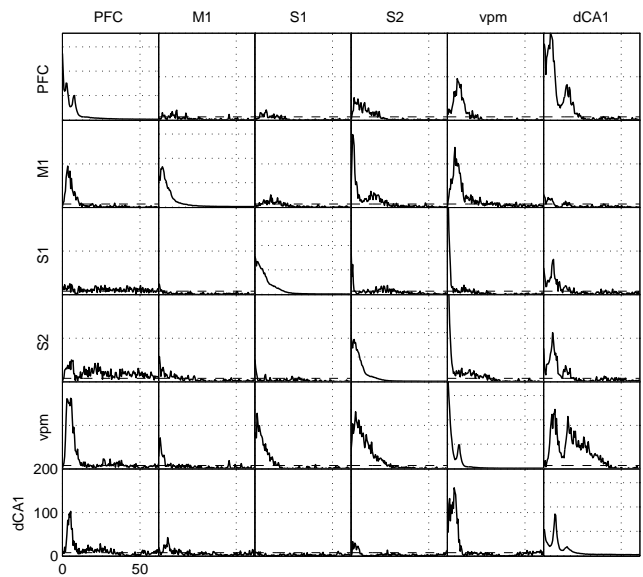}

\caption{Renormalized PDC between several part of the brain in the sleeping mice experiencing paradoxical sleep. The dashed line correspond to the Bonferroni  threshold for a Family Wise Eror Rate of 0.05.  All plots are on the same scale as depicted in the lower left plot. For all the plots, the $x$ axis is labelled in frequency from 0 to 62.5 hz.  }
\label{ExempleReel:fig}       
\end{figure}

The results are presented in figure (\ref{ExempleReel:fig}).
Since we do not want to draw definitive conclusion regarding neuroscience (this would require much more analysis, a better statistical analysis in terms of animals recorded, {\it etc}), we just analyse some features revealed by the analysis. 
First, we must come back to the discussion of the physical meaning of the analysis. In terms of Granger causality, the fact that the renormalized PDC overpass the significance threshold at some frequency means that one signal is a cause of the other,  given the set of observation. It does not give any information on the physical reality linking the two signals. 
 If we interpret the result as an energy flow from one area to the other, we must use the result with the caution recalled earlier. This represent only a linear modeling of the links, and the renormalized PDC in a given frequency bin overpassing the significance level only reveal that their may be some linearity in the link between two areas.
 
 The first striking feature is the high dissymmetry in the links. For example, dCA1 causes S1 but S1 does not cause dCA1. dCA1 causes 
 all the other areas except M1, since the corresponding PDC in very comparable to the threshold. On the contrary only vpm and PFC (essentially) causes dCA1. M1 is not a cause of almost all the other areas, but is caused mainly by PFC, S2 and vpm. 
 As mentionned earlier, we do not go further in the interpretation in this paper since it is not its the goal. Work on the use of the method explained here on neuroscience experiments is ongoing and will appear later. 
 
\section{Discussion}

The main contribution of this paper is the use of a clever algorithm of spectral factorization as a trick to identify
causal filters between different time series. The full procedure relies on the estimation of a spectral density matrix using usual tools of times series analysis, and the application to the inverse of this matrix of the Davis\&Dickinson algorithm for spectral factorization.  
The  spectral factors thus obtained reflect the direct link between pairs of signals. They can be used in the calculation of well known measures in neuroscience such as the Partial Directed Coherence, or its renormalized version which is easier to use in practical testing. 

The contribution is therefore essentially algorithmic. But before discussing the advantages and drawbacks of the method and its comparison to others, we want to insist on an important point of the method: it is inherently non parametric. When dealing with Granger causality, this is especially important because in essence, testing  Granger causality between two times series
amounts to testing nullity of the transfer between them, or equivalently, testing nullity of the corresponding entry in the spectral factor matrix. This comes from the invertibility of Wold decomposition, as recalled earlier in this paper, but as previously stated by \cite{Gewe82,Gewe84}, and recalled later by {\it e.g. }   \cite{Eich06}. And indeed, Granger causality is a nonparametric concept.

\paragraph{Comparison with AR modeling.}

To begin with, let us recall the usual way for calculating the PDC or its renormalized version. The method relies in identifying the autoregressive model that we recall here
\begin{eqnarray*}
\vx(t) = \sum_{k=1}^{q}\vA(k) \vx(t-k) + \vvarepsilon(t) 
\end{eqnarray*}
that is estimating the matrices $\vA(k), k=1,\ldots,q$ and the covariance matrix $\vSigma_{\varepsilon}$ of the zero mean i.i.d. noise $\vvarepsilon(t) $. Practically, the identification uses the least mean square algorithm: using $KN$ observations $ \vx(1), \ldots, \vx(KN)$ (to compare with the method developed here, the time over which we learn is taken the same and is $KN$), we have to estimate the matrices. In order to do this, 
construct the matrices $\vY = ( \vx(1), \ldots, \vx(KN) )$ of size 
$p \times KN $, $\vX$ of size $(pq) \times KN $ where the $t$-th column is $(\vx(t-1)^\top ,\ldots, \vx(t-q)^\top)^\top$, and $\vE  = (\vvarepsilon (1),\ldots, \vvarepsilon (KN) $ of size  $p \times KN $. The matrices $\vA(k)$ are stored in the  $p \times (q p)$ matrix $\vB= (\vA(1),\ldots, \vA(q))$.
Then we get the compact  equation $\vY = \vB \vX +\vE$. Then the least square solution is given by $\widehat{\vB}= \vY \vX^\top (\vX \vX^\top)^{-1}$.
From this estimate an estimate of $\vSigma_{\varepsilon}$ is obtained as $\widehat{\vSigma_{\varepsilon}}= (\vY-\widehat{\vB}\vX)(\vY-\widehat{\vB}\vX)^\top/(KN)$. The spectral factors are then obtained by Fourier transforming the corresponding estimated impulse responses
contained in $\widehat{\vB}$.

We can now turn to a complexity analysis. We begin with the autoregressive approach. The matrix to invert costs $O( pq KN) $ to be built, whereas its inversion costs $O( p^3q^3 )$. The product $\vY \vX^\top$ costs $O( p KN)$ where the last product costs $O(p^2 q)$. 
Thus overall, the ordinary least square identification costs $O( p^3q^3  + pq KN)$. 

For a $p$ dimensional process cut into $K$ block of length $N$ samples, the computational complexity for the spectral factorization approach is as follows.  In the estimation procedure, we perform $p$ FFT of length $N$ at a cost of $O(N \log N)$ multiplications for each. The $p$ FFT obtained are used to created the matrix of periodograms, and this costs $O(p^2N)$ multiplications
This is done $K$ times and the total cost is $O(p^2 K N + pKN\log N) $. The matrix inversion has to be done for the $N$ frequency for a total 
cost of $O(p^3 N)$. In the spectral factorization algorithm, we have to invert a matrix at a cost of $p^3$, make four multiplications of square  matrices of size $p$ at every frequency for a cost of $O(Np^2)$, apply twice the FFT for $p^2$ signals, and multiply in between by a vector of size $N$ each (causal projection) for a total cost main cost of $O(p^2 N \log N)$. The test for stopping costs $O(N p^2)$. Since the number of iteration of the algorithm is in general limited (typically from tens to some tens), the spectral factorization costs $O(Np^3 + p^2N\log N)$. 
Thus overall, for reasonable dimensions, the evaluation of the spectral density is the most costly for $O(pKN\log N)$.

Obviously, if $p$ and $q$ are small (compared to $KN$), then the cost of the autoregressive identification is $O(pq KN)$, better than 
 $O(pKN\log N)$, but only slightly better since $\log N$ is far from being big! However, suppose that the order $q$ is found to be of the same order as $N$. Then the cost of the autoregressive identification is  $O( p^3N^3  + p KN^2)$ which is more than one order of magnitude 
 higher than  $O(pKN\log N)$. 
 
Therefore, in terms of complexity, the proposed method is comparable to the usual method if moderate orders are required, but is far more rapid in the case of high autoregressive orders. Note that in term of $p$ the complexities of both are comparable.

 One of the drawback for the moment is the absence of explicit form for the statistics of the estimated spectral factors, even asymptotically.
 However, we  conjecture that these are asymptotically unbiased and complex normally distributed, as obtained by smooth transformation of asymptotically Gaussian random variables (invoking the delta-method). This fact was verified on simulation but remains to be proved. 
 Further, the covariance of the estimates is unknown, even in the asymptotic case, contrary to the autoregressive approach. Thus, in order to normalize appropriately the PDC, we have recourse to a bootstrap approach which obviously requires an effort in computation time.

\section{Appendices}

\subsection{Spectral factorization}
\label{algoDickinson:app}

The aim here is to present the main steps for the derivation of Davis\&Dickinson algorithm. A  code is presented in the following appendix. The algorithm relies on the equivalence between Kalman filtering and Wiener filtering.
Note that the complete proof is lengthy and requires a lot of algebraic manipulation. The complete proof is given in some detailed in Dickinson's master thesis, but does not appear in other publications. This is is the main reason to include here the main steps of this proof.
The only difference with Dinckinson's proof is the faster way we use to obtain eq. (\ref{interm:eq}) below.

The proof consists in expressing the spectral factors used in Wiener filter in terms of the elements of the solution of Kalman filtering. Then the spectral factor essentially depends on the covariance of the error which is given by the solution of a Riccati equation. This equation has no closed form solution (except in very rare cases). Using a Newton-Raphson recursion to solve the Riccati equation allows as a by product to obtain Davis algorithm.

Suppose we have the following  state and observation  equations
\begin{eqnarray*}
\vx_k&=& \vA \vx_{k-1}  + \vB \vu_k \\
\vy_k &=& \vC \vx_k  + \vv_k
\end{eqnarray*}
where $\vu$ and $\vv$ are independent white sequences with zero mean and respective covariance matrices $\vQ$ and $\vR$.
The aim in filtering is to estimate $\vx_k$ from the observation $\vy$ up to time $k$.

The covariance matrix is $\vS_{yy}(z) = \vC \vS_{xx}(z) \vC^\top + \vR$ and can be written 
\begin{eqnarray*}
\vS_{yy}(z) = \vR + \vC  (\vI z - \vA)^{-1} \vB \vQ \vB^\top (\vI z^{-\star} - \vA)^{-\dagger}\vC^\top
\end{eqnarray*}
The Wiener filter necessitates to have a strong spectral factorization of this spectral matrix in the form
\begin{eqnarray*}
\vS_{yy}(z)  = \vF(z)  \vW  \vF^{\dagger}(z^{-\star})
\end{eqnarray*}
Then the spectral factor is given by $\vF(z) = \vI + \vC (\vI z-\vA)^{-1} \vK$ where the Kalman gain (steady state)
reads $\vK = \vA \vP \vC^\top \vW^{-1}$ and $\vW = \vR + \vC \vP \vC^\top$. $\vP$ is the solution of the Riccati equation
$\vP = \vA \vP \vA^\top - \vA \vP \vC^\top \vW^{-1} \vC \vP \vA^\top + \vB \vQ \vB^\top$.

The problem  reduces to obtain the solution of the Riccati equation, which is far from being obvious.
For this, Davis proposed to use a Newton-Raphson algorithm for solving
\begin{eqnarray*}
0&=& f(\vP) \\ 
&=& -\vP + \vA \vP \vA^\top - \vA \vP \vC^\top \vW^{-1} \vC \vP \vA^\top + \vB \vQ \vB^\top
\end{eqnarray*}
 The Newton-Raphson iteration for solving this is
 \begin{eqnarray*}
\vP_{n+1} - (\vA - \vK_n \vC) \vP_{n+1} (\vA - \vK_n \vC)^\top =\\ \vK_n \vR \vK_n^\top + \vB \vQ \vB^\top
\end{eqnarray*}
A first trick is to use the the representation in series  $\vX=\sum_{n\leq0} \vA^{n} \vGamma \vA^{\top n }$ 
for  the solution of $\vX+ \vA\vX\vA^\top=\vGamma$. If $\vGamma$ is positive definite, the series is an inner product, and we can use
Parseval equality to obtain the equivalent form in the $z$ domain.
Apply this to $\vP_{n+1}$ to obtain
\begin{eqnarray*}
\vP_{n+1}  &=& \frac{1}{2i\pi}  \oint  (z\vI - \vA+ \vK_n \vC)^{-1} \big( \vK_n \vR \vK_n^\top + \vB \vQ \vB^\top\big) \\
& &(z^{-\star} \vI - \vA+ \vK_n \vC)^{-\dagger} \frac{dz}{z}
\end{eqnarray*}
Then pre- and  post-multiplying $\vP_{n+1}$ by $\vC$, using some algebra and remembering the definitions of $\vS_{yy}$ and $\vF$ and the fact that $(2i\pi)^{-1}\oint \vF_n(z)^{-1} dz/z=\vI$ allows to obtain 
\begin{eqnarray*}
\vR+\vC \vP_{n+1} \vC^\top &= & \frac{1}{2i\pi}  \oint  \vF_n(z)^{-1} \vS_{yy}(z)  \vF_n(z^{-\star})^{-\dagger}\frac{dz}{z} \\
&:=& \vW_n
\end{eqnarray*}
which constitute the first part of the algorithm.

To get the iteration on the spectral factor, Dickinson proposes to study $\Delta \vP_n := \vP_{n+1}-\vP_n$. 
Substracting two successive iteration of the Newton-Raphson iteration leads to 
\begin{eqnarray*}
\Delta \vP_{n} - (\vA - \vK_n \vC) \Delta \vP_n (\vA - \vK_n \vC)^\top  = \\-(\vK_n- \vK_{n-1})\vW_{n-1}(\vK_n- \vK_{n-1})^\top 
:=- \vT_n
\end{eqnarray*}
This last matrix $ \vT_n$ is positive definite since $\vW_{n-1} $ is positive definite. 

Since $\Delta \vP_{n}$ satisfies an equation of the type   $\vX+ \vA\vX\vA^\top=\vGamma$, we use the series representation for 
 $\Delta \vP_{n}$, 
 \begin{eqnarray*}
\Delta \vP_{n} = -\sum_{k\geq 0} (\vA - \vK_n \vC)^k \vT_n(\vA - \vK_n \vC)^{(k)\top}
\end{eqnarray*}
and since $ \vT_n$ is positive definite, we can have an equivalent form in the $z$ domain
\begin{eqnarray}
 \Delta \vP_{n} &=&\frac{-1}{2i\pi}  \oint  \big(\vI - z^{-1}(\vA- \vK_n \vC)\big)^{-1}\vT_n  \nonumber\\
 & &\Big(z^{-\star}  ( \vI - z^{\star}(\vA- \vK_n \vC)\Big)^{-\dagger} \frac{dz}{z} 
 \label{interm:eq}
 \end{eqnarray}

Then we have to solve equation which can be verified by direct evaluation
\begin{eqnarray*}
 \vC(z \vI - \vA)^{-1}(\vA-\vK_n \vC)^\top \Delta \vP_n  \vC^\top 
= (\vF_{n+1}(z) -\vF_n(z)) \vW_{n}\\
\end{eqnarray*}
Inserting (\ref{interm:eq}), a lengthy calculation leads to 
 \begin{eqnarray}
&& -2i\pi (\vF_{n+1}(z) -\vF_n(z)) \vW_{n}  \\
&=& (\vF_n(z)- \vF_{n-1}(z)) \vW_{n-1} \nonumber   \\ 
&\times & \oint  \frac{dv}{(v-z)} ( \vF_n(v^{-\star})-  \vF_{n-1}(v^{-\star}))^{\dagger} \vF_n(v^{-\star})^{-\dagger}\nonumber \\
&-& \vF_n(z) \nonumber \\
&\times&\oint  \frac{dv}{(v-z)} \vF_n(v)^{-1}  \Big( \vF_n(v) \vW_{n-1} \vF_n(v^{-\star}) -\vS_{yy}(v)  \Big)\vF_n(v^{-\star})^{-\dagger} \nonumber 
\label{final:eq}
\end{eqnarray}
an  expression which can be linked with the causal projection.

If $H(z)$ is the $z$ transform of a sequence $h_k, k\in \Z$, remember that 
\begin{eqnarray*}
(P_+H)(z) &= &\sum_{k\geq 0} h_k z^{-k}\\
&=&\frac{1}{2i\pi }\oint  \frac{dv}{v} \frac{z}{z-v} H(v)
\end{eqnarray*}
Thus, note that the integrals appearing in expression (\ref{final:eq}) are of the form
\begin{eqnarray*}
\oint  \frac{dv}{v} \frac{v}{v-z} H(v) &=& \oint  \frac{dv}{v} \frac{v+z-z}{v-z}H(v) \\
&=& \oint  \frac{dv}{v} H(v) -2i\pi (P_+H)(z)
\end{eqnarray*}
The first integral in   (\ref{final:eq})  concerns an anticausal quantity with no constant term and is therefore equal to zero. Thus we have 
Noting that $\vW_{n-1}$ does not depend on $v$ and therefore its causal part is equal to itself,  we finally get the beautiful result
\begin{eqnarray*}
\vW_{n}  &=& \frac{1}{2i\pi }\oint \frac{dv}{v}\vF_n(v)^{-1}\vS_{yy}(v)\vF_n(v^{-\star})^{-\dagger}\\
\vF_{n+1}(z) &= & \vF_n(z) \Big(P_+\big[  \vF_n^{-1}\vS_{yy}\vF_n^{-\dagger}\big]\Big)(z)  \vW_{n}^{-1}
\end{eqnarray*}

\section{Code for spectral factorization}
\label{Code:app}

The following is a Matlab$^\copyright$  code for the spectral factorization. 
It uses three dimensional arrays. No test for positive definiteness is included, and if the assumption
on the matrix $S$ are not adequate, the algorithm should not converge.

{\small
    \begin{verbatim}
    
function [F,W]=spectral_factorization(S)

% Provide the strong spectral factorization
% S= F W F^h of the spectral matrix S.
%
% S : dimension n*n*mf is the spectral matrix
% F : dimension n*n*mf is the spectral factor
% W : dimension n*n are the weights
%
% PO Amblard 2013
% based on Davis&Dickinson algorithm, 
% SIAM J. Appl. Math, 43, 2, pp 289--301, 1983
%
% [F,W]=spectral_factorization(S);

[n m mf]=size(S);
F=zeros(n,n,mf);W=zeros(n,n);G=F;GC=G; err=zeros(1,mf);
Ustep=[1 ones(1,mf/2-1) 1/2  zeros(1,mf/2-1)];

% initialize F to identity for all frequencies
for f=1:mf;  F(:,:,f)=eye(n); end; FI=F;

% iterations.
tol=1.0000e-06;  % tolerance could be passed as a parameter.
err_new=1;
while (err_new>tol)
    for f=1:mf;
       G(:,:,f)= FI(:,:,f)'*S(:,:,f)*FI(:,:,f);
    end
    W=real(mean(G,3));WI=W\eye(n);
    for i=1:n;
        for j=1:n;
            GC(i,j,:)=fft(ifft(squeeze(G(i,j,:))).'.*Ustep);
        end
    end
    for f=1:mf;
        F(:,:,f)=WI*GC(:,:,f)/2*F(:,:,f);
        Sest(:,:,f)=F(:,:,f)'*W*F(:,:,f);
        err(f)=norm(Sest(:,:,f)-S(:,:,f),inf);
    end
    err_new=max(err);
end
\end{verbatim}}

{\bf Acknowlegments:} P.O. Amblard is supported by a Marie Curie International Outgoing Fellowship of the European Union. P.O. A. gratefully acknowledges J. Davis for the discussions on his algorithm, and S. Crochet for making available his data. 

\bibliographystyle{unsrt}      

\end{document}